\documentclass{article}

\usepackage{arxiv}

\usepackage[utf8]{inputenc} 
\usepackage[T1]{fontenc}    
\usepackage{url}            
\usepackage{booktabs}       
\usepackage{amsfonts}       
\usepackage{nicefrac}       
\usepackage{microtype}      
\usepackage{graphicx}

\usepackage{hyperref}
\usepackage{doi}

\title{A Generalised Logical Layered Architecture for Blockchain Technology}

\author{Jared Newell*,
        Quazi Mamun*,
        Sabih ur Rehman*
        and Md Zahidul Islam*\\ 
        {*School of Computing, Mathematics and Engineering, Charles Sturt University, Australia}}

\begin{document}
\maketitle

\begin{abstract}
	Precision, validity, reliability, timeliness, availability, and granularity are the desired characteristics for data and information systems. However due to the desired trait of data mutability, information systems have inherently lacked the ability to enforce data integrity without governance. A resolution to this challenge has emerged in the shape of blockchain architecture, which ensures immutability of stored information, whilst remaining in an online state. Blockchain technology achieves this through the serial attachment of set-sized parcels of data called blocks. Links (liken to a chain) between these blocks are implemented using a cryptographic seal created using mathematical functions on the data inside the blocks. Practical implementations of blockchain vary by different components, concepts, and terminologies. Researchers proposed various architectural models using different layers to implement blockchain technologies. In this paper, we investigated those layered architectures for different use cases. We identified essential layers and components for a generalised blockchain architecture. We present a novel three-tiered storage model for the purpose of logically defining and categorising blockchain as a storage technology. We envision that this generalised model will be used as a guide when referencing and building any blockchain storage solution.
\end{abstract}


\section{Introduction}
Throughout the history of civilisation, record keeping has been necessary to maintain the integrity of transactions and covenants between parties, governmental structures, and autonomous civilisations \cite{hudson2018}. The oldest existing immutable record storage, known as the cuneiform clay tablets, dates from around 3500 BC. It contains mostly debt contracts and administrative accounts, and demonstrates some of the earliest methods for immutable storage. Immutable storage has continued to function, mostly in the form of paper, until only recently when there has been a significant decline in its use within society. With the diminishing cost of digital storage, and expanding capabilities of geographically distributed ‘cloud’ storage platforms, immutability of information has become one of the most challenging problems \cite{Goda2012}. In addition to the immutability of data and information, other essential and desirable qualities of data and information need to be maintained. These include accuracy or precision, legitimacy or validation, reliability and consistency, timeliness and relevance, availability or accessibility. Consistency and integrity of the information are maintained through the use of access control mechanisms, policies, authorisation by both administrative and technical controls \cite{yaokumah2017}. In implementing the qualities mentioned above, information systems have relied on quasi-immutable storage solutions known as offline or cold storage, in which information is copied at a specific point in time. A criticism of offline storage is that it lacks the ability to readily audit the stored information. 
Moreover, computer storage from its inception has been unable to maintain integrity without governance, with the possible exception of punched card storage \cite{gorn1966}. The onus of record integrity and accuracy is left to the custodian of the information at times of conflict which is the main criticism of read-write computer storage \cite{hoffman1969}. A solution to this problem is the blockchain technology. Blockchain is a software technology that uses traditional online digital media and cryptography to ensure the integrity of the information being stored. This is done by transforming online digital storage into a write once, read many (WORM) storage technology, and thus it introduces immutability for the information being stored on it. Blockchain can maintain integrity without governance, and this is what makes this online storage technology unique in its design. However, it lacks a unified definition of what blockchain is and how it is to be used when including it in a solution. There are no standards for the core components, generalised architecture, and associated components of a  blockchain. Defining blockchain’s essential components will help to build a universal classification and can identify blockchain as an immutable storage technology that introduces the ability to form compatibility between designs.
In doing so, in this paper we studied different use cases of blockchain, identified the different layered architectures used in those use cases. We also identified different components of the blockchains in different use cases. We then associated the essential components to the identified layers of a generalised blockchain architecture. 

\begin{figure}[t]
    \begin{center}
    \includegraphics[width=0.75\columnwidth]{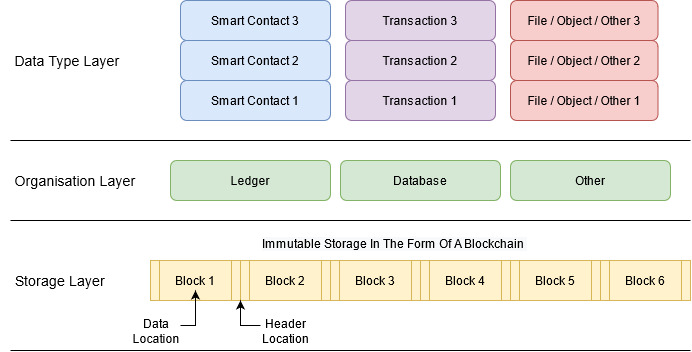}
    \caption{{The logical layered structure of blockchain implementation.}}
    \label{fig:blockchainstoragecomponents}
    \end{center}
\end{figure}

This paper begins with an investigation of the origins of blockchain and its supporting technologies in section \ref{thebeginning}. Section \ref{blockchainasinformationstorage} continues with the identification of the generalised mechanisms behind the workings of a blockchain. Section \ref{theblockchainstoragelayermodel} examines the functionalities of the blockchain technology. In section  \ref{relatedworkusingblockchainasstorage}, we investigated different architectures used to implement blockchain for different use cases. In doing so, the use cases were categorised by industry, field or technology, and we identified the logical layers used in each group for blockchain implementation. In section \ref{blockchaincomponetsdefined} we define and categorise the blockchain technologies using our three-tiered model, starting with the core blockchain components of  the storage layer, and the associated blockchain components of the organisation layer and the data type layer. The paper further examines some future considerations related to the classification of blockchain and the function of the storage layer model in section \ref{futureconsiderations}. Finally, we conclude by highlighting the contributions of the blockchain storage model to the identification and categorisation of the blockchain core technologies and its associated technologies to aid in the architecture of future immutable storage solutions in section \ref{conclusion}.

\begin{figure}[t]
    \begin{center}
    \includegraphics[width=0.75 \columnwidth]{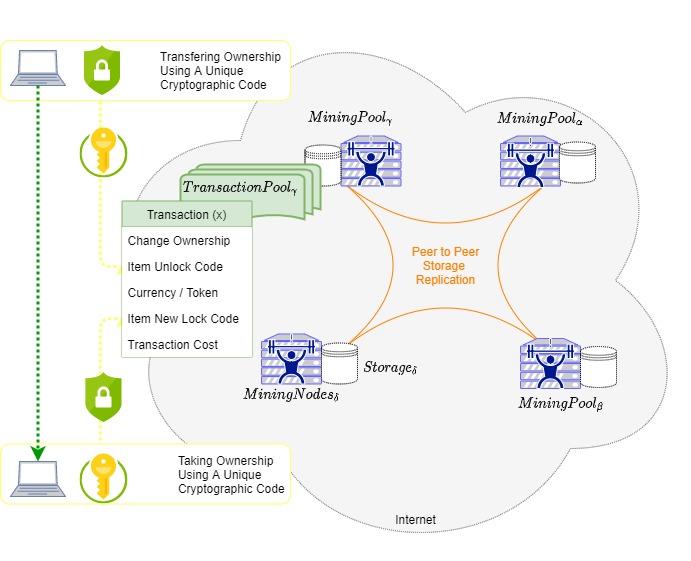}
    \caption{{The Bitcoin Blockchain Network - Two clients executing a transaction.}}
    \label{fig:blocktransaction}
    \end{center}
\end{figure}
 \section{The Beginning\label{thebeginning}}
 The first use case of blockchain technology can be identified as the use case of a cryptocurrency called Bitcoin. In 2009, Satoshi Nakamoto released a functional solution for a decentralised digital currency; a suite of software technologies that included transaction processing and storage replication protocols, conflict resolution in the network and controls for a new digital currency \cite{nakamoto2008}. A novel technology called blockchain was included in this suite of software, presented in the storage layer of figure \ref{fig:blockchainstoragecomponents}, and was fundamental to the success of the digital currency or cryptocurrency called Bitcoin. Satoshi Nakamoto expanded on some of the previous attempts to develop a digital currency – these included Hashcash, Bit Gold and X-Cash (executable cash), with some of the previously mentioned being proposed ten years earlier, in \cite{claessens2003}, and B-Money only a few months before the release of Bitcoin. Satoshi Nakamoto’s solution included the currency’s ledger of account, which is responsible for tracking each piece of cryptocurrency throughout its lifecycle. This ledger is stored on a blockchain, existing in the organisation layer in figure \ref{fig:blockchainstoragecomponents} and is then distributed, as copies, throughout its network. The ownership of Bitcoin within the ledger is determined by a private key, which matches an associated public key on the blockchain, described in \cite{diffie1976}. This process is represented by the transaction displayed on the left of figure \ref{fig:blocktransaction}. This keypair is mathematically linked using keypair cryptography, described in \cite{Diffie1988}, to the owner’s virtual wallet, using elliptic curve cryptography \cite{koblitz1987}. The wallet exists as a software or hardware solution and produces an address used to transact. All transactions are valid once they are recorded on the blockchain.

\begin{figure}[t]
    \begin{center}
    \includegraphics[width=0.75 \columnwidth]{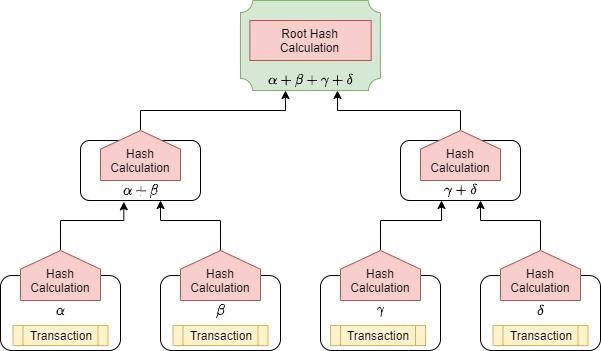}
    \caption{{Transaction validation using a binary Merkle tree.}}
    \label{fig:binarymerkletree}
    \end{center}
\end{figure}

\begin{figure}[t]
    \begin{center}
    \includegraphics[width=0.75 \columnwidth]{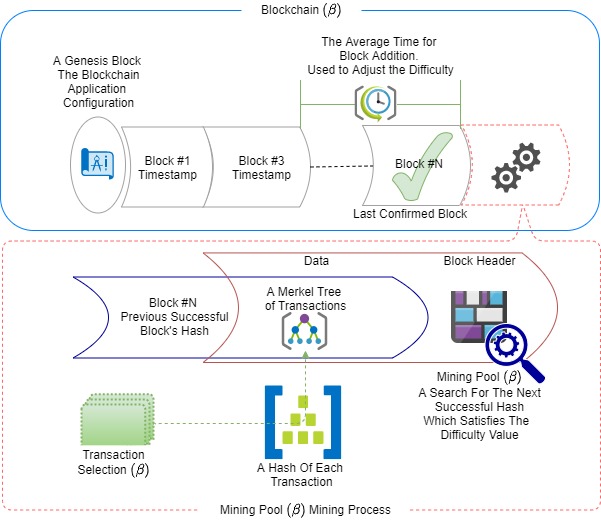}
    \caption{{The Bitcoin blockchain – A potential block and its components during the mining process.}}
    \label{fig:blockchainminingmechanics}
    \end{center}
\end{figure}

The Bitcoin blockchain design is described as linking data blocks, with the data being transaction updates to the ledger. The key attributes of Bitcoins blockchain are presented in figure \ref{fig:blockchainminingmechanics}, located in the storage layer of figure \ref{fig:blockchainstoragecomponents}. This novel approach to storage formed the first known, practical online immutable storage technology. Within this architecture, a binary Merkle tree of hashes, visualised in figure \ref{fig:binarymerkletree}, represents all the transactions processed and included in the data location of the block, presented in figure \ref{fig:blockchainminingmechanics}. The Merkle tree is used to verify that a recorded transaction remains unchanged, as it is only possible for a single hash to represent a single piece of data \cite{merkle1980}. The root hash at the top of the tree, and other header parameters, is used to build the mathematical seal. This, in turn, is used in the seal of the next block in the blockchain, displayed in the lower half of figure \ref{fig:blockchainminingmechanics}.

The Bitcoin blockchain routines are executed using a support network of computers or nodes, which provide the storage for the blockchain, the replication of the blockchain blocks, and the verification of the ledger – displayed within the cloud in figure \ref{fig:blocktransaction}. One of the nodes in the network will add the next block to the chain – this requires generating a sealing hash as a mathematical proof. This hash must meet specific requirements, controlled by the adjusting difficulty parameter at the top of figure \ref{fig:blockchainminingmechanics} and because of this, it takes a significant number of attempts, resources and time to complete. This process is known as mining in Bitcoin. A visual representation of a block being added into a Bitcoin blockchain is presented at the bottom of figure \ref{fig:blockchainminingmechanics}.
A decentralised Peer-to-Peer (P2P) networking protocol is used for replication of state changes between the distributed nodes \cite{parameswaran2001} is displayed in figure \ref{fig:blocktransaction}. All coordination within the distributed, decentralised, transaction network is maintained with a consensus algorithm. In Bitcoin, it is called Proof-of-Work (PoW), formalised in \cite{jakobsson1999}. In Bitcoin, the Secure Hash Algorithm 256 is used, explored in \cite{perez2018}, for the Merkle tree, the mathematical seal for each block \cite{zohar2015} and aspects of the ledger’s contents. This hashing function is considered computationally expensive. In addition, Satoshi Nakamoto designed currency controls implemented within the software, such as the inability to release more currency than defined within a set period and to reduce the amount released after a set interval \cite{bohme2015}. These controls are expressed statically in the Genesis block, the first block in the blockchain and the dynamic configuration of the average time for block addition, presented in the top of figure \ref{fig:blockchainminingmechanics}. After 11 years of continuous online operation, the Bitcoin blockchain is the longest-running example of a functional blockchain using this architecture \cite{yli2016}.

It must be highlighted that the terminology used and concepts discussed can be interchangeable with other blockchain architectures, however, they do not necessary correlate in their characteristics or application. The blockchain technology, at the time of writing, remains fundamentally application specific.

\section{Generalised Mechanisms Behind Blockchain} \label{blockchainasinformationstorage}
In this section we seek to define a generalised description, characteristics and mechanisms universally adopted throughout blockchain solutions. The generalised function of blockchain is a method for digital computer storage, which exists as an additional layer on top of current computer storage technologies. The blockchain storage method consists of two phases; first, the blockchain adopts unidirectional cryptogram to store a representation of the data in the form of a message digest and the original raw data. As mentioned in \cite{jin2018} and \cite{li2018}, in some implementations of blockchain, the raw data is stored encrypted for data privacy. Encryption in this process is unrelated to the underlining blockchain storage core functions, as it is applied to the data being stored. In the second phase, the blockchain uses a cryptographic process that produces a proof; a mathematically verifiable calculation that the block and its associated block directly adjacent have remained unchanged \cite{yli2016}. Once the first block is established, known as the genesis block, these two phases are repeated ad infinitum, forming an online, digital, immutable storage solution in the form of a chain of blocks.

Due to the blocks being serially attached, they are therefore chronologically ordered. The immutability of the information stored is enforced because any change in a previously stored block would require the rewrite of each block after that. Furthermore, this includes calculating each of the associated proofs and a plausible timestamp, while simultaneously being accepted by the network participants. Moreover, as the chain increases in block count, the likelihood of a successful modification reduces \cite{zhang2019}.

The aforementioned fundamentals of blockchains remain the same. The generalised mechanism of a blockchain storage technology is confined to the core component in the storage layer of the three-tier model defined in figure \ref{fig:blockchainstoragecomponents}. At the storage layer in this figure, individual blocks are segregated in two parts, first is the data location used for storage and the second is the header location which contains parameters required for the chaining together of the blocks. Blockchain implementations have varied with time – best generalised by the characteristic of a compromise between the rate at which data can be stored and the complexity in mathematical difficulty required to link the blocks together, also expressed as the security of the blockchain.

\section{The Blockchain Layer Model} \label{theblockchainstoragelayermodel}

Blockchain has been identified as a key storage solution within a number of different fields and associated technology solutions \cite{yli2016} \cite{casino2019}. However, standadisation of blockchain lacks the clarification, classification and categorisation of which technologies make up core elements of blockchian technology and which can be separated as independent technologies attributed to blockchain technologies. This is especially apparent when reviewing related works about blockchain, highlighted in section \ref{relatedworkusingblockchainasstorage} and asserted in \cite{risius2017} and \cite{koens2018}. It must be noted that attempts have been made, such as in \cite{butijn2020}, to define the core blockchain properties, with the need to clarify blockchain terminology raised in \cite{halaburda2018}. We analyse existing blockchain models and use these to compare and present our novel three-tiered storage model, which uniquely describes blockchain from the aspect of a storage solution, the solution being a method for immutable online computer storage.

We begin with the identification and comparison of a selection of existing blockchain models. In \cite{Yuan2018} and \cite{Xinyi2018} a six-layer model is presented consisting of data, network, consensus, incentive, contract, and application layer. In reviewing this model we reflect on what we consider core blockchain components that are required to build an immutable storage solution and which are not. From the six-layer model we combine all the components from the data, network and consensus layers into a single layer called the storage layer, we abstract this further into core component groups required to build a blockchain solution, discussed in details in section  \ref{blockchaincomponetsdefined}. However, we exclude the incentive layer of the six-layer model as it is used only in cryptocurrencies or trading specific blockchain implementations. The authors also acknowledge this as an optional layer in \cite{Yuan2018}. In \cite{Duan2018} the authors present a five-layer model excluding the incentive layer. Furthermore, the contract layer of the above models are not implemented universally in blockchain designs and is therefore excluded as a core blockchain component. However, we present this as an associated technology at the data type layer in our model in figure \ref{fig:blockchainstoragecomponents}. Finally, the application layer of both the six-layer model and the five-layer models list industries and generalised applications which make use of the information stored on the blockchain. In our model we remove this, as a single blockchain design can satisfy the requirements in more than one industry or technology solution. In addition we do not consider it contains any core blockchain component required to function as an immutable storage solution, but when to use a blockchain. In \cite{Clavin2020} the authors presents a 3 layer blockchain  model, as per the models above we exclude the application layer, and smart contracts. We consider the consensus layer included as a core component at the storage layer of our model, however, there is a number of blockchain components which have not been considered and is therefore lacking in adequately presenting all components and considerations of blockchain. In \cite{ellervee2017} a reference model is presented with two sections of roles and actors, and of services and processes. The services and processes section consist of the blockchain components of transaction, network discovery, consensus and block generation. Within the services the components are highlighted if it is used only in permissioned blockchains. Within this model we include all components associated with block generation at the storage layer of our model. However, we include the transaction related components as not core blockchain components but associated blockchain technologies. A transaction is the type of data being stored and exist within the data section of each block. Moreover, a transaction becomes immutable once stored on the blockchain but is not required to create an immutable storage solution, presented in figure \ref{fig:blockchainstoragecomponents}. We consider our  model not to conflict but to limit the scope of components to a generalised subset of that which is required to design, reference and build blockchain.

\ref{fig:blockchainstoragecomponents} depicts the proposed logical model using three distinct layers. We discuss the purpose for each layer first with the storage layer. This layer consists of the core blockchain components required to obtain the immutability of online digital data storage. The characteristics of the storage layer consist of properties such as data immutability, chronological order, security, mathematical proofs and cryptography representing the blockchain as an immutable data storage technology. With the addition properties of its supporting network that provide the data write ability for a blockchain: being a consensus, computer hardware, fault-tolerance, controls, scalability, centralisation, decentralisation, replication, permission and accessibility. This writing  process is through block addition this is represented by the mining nodes in the Internet cloud of figure \ref{fig:blocktransaction}.

Second and placed above the storage layer is the organisation layer presented in figure \ref{fig:blockchainstoragecomponents}. Its purpose is to decouple the association of the blockchain, as an immutable storage technology, with the structure or organisation of the information being stored, and thus forming the relationship between all the data locations across all blocks in the blockchain. The data location in each block is presented in figure \ref{fig:blockchainstoragecomponents}, alongside the block header location, within the storage layer, with additional detail of the data location in figure \ref{fig:blockchainminingmechanics}. The storage structure in the organisation layer often takes the form of a transaction ledger – when the blockchain is replicated to more than one node, this then adopts the name of a distributed ledger. 

This model's last layer is the data type layer, which decouples the data structures such as a single transaction or smart contract, which is a computer code stored on the blockchain, presented in figure \ref{fig:blockchainstoragecomponents}. This layer's data type often has a close relationship with the structure used in the organisation layer. For example, a transaction is often associated with a distributed ledger. However, transaction types can be customised to fit the application of the blockchain. In this model, referred to in figure \ref{fig:blockchainstoragecomponents}, a focus has been placed on the blockchain core and the associated technologies within the context of storage. Therefore, it excludes functions such as the execution of the smart contracts, as this is performed by external compute resources, documented in \cite{wang2019}. We presented the novel three-tiered layer model that provides a logical structure that can be used to differentiate technologies and terminologies that are blockchain-related to categorise, classify and define blockchain.

\section{Blockchain Implementation} \label{relatedworkusingblockchainasstorage}

We combine the use of our blockchain layer model with a limited selection of reviewed literature, selected from papers where blockchain is used in the title. The purpose is to evaluate how blockchain is implemented and referenced in each solution and is then tabulated. This table reflects our three-tiered layer model from figure \ref{fig:blockchainstoragecomponents} with the columns of storage layer, organisation layer, data type layer presented in table \ref{tab:StorageLayers}, highlighting what each piece of literature implements the layers. We use this to reinforce the relevance of our model in defining blockchain. The literature selected is grouped by industry, field or technology in the following sections.

\textbf{\textit{The Finance industry}} - Blockchain’s purpose as a distributed ledger for cryptocurrencies' financial transactions has extended into traditional financial transactions. In \cite{hyvarinen2017}, blockchain is used as a ledger to represent financial transactions as tokens. This is to prevent tax fraud by replacing the current system and removing double claiming of tax refund amounts. The author lacks specifics on how the blockchain is implemented, however, it can be inferred that a distributed ledger organisation structure using a smart contract data type is used to automate the execution of verification tasks in the solution. In \cite{dai2017}, a permissioned blockchain is combined with enterprise resource planning software. This is to produce a triple-entry accounting solution by using a ledger organisation structure in addition to the automation of business processes using a smart contract data type as the solution.

\textbf{\textit{Energy sector}} -  \cite{andoni2019} review the current use and potential uses of blockchain to create new markets in energy production and the associated carbon trading. The distributed ledger data structure is used, incorporating the use of the data type of smart contracts. The overall solution is to reduce overheads or expenses by automation of administrative tasks and trading, especially when the energy market is decentralised and at small scale. In \cite{chen2020}, blockchain is used to the collect of data from smart grids, generally for billing purposes. A permissioned, parallel, blockchain implementation is constructed, focusing on the transaction processing rate, and the authentication and privacy of its transactions.

\begin{table}[t]
\begin{center}
\centering
\scalebox{0.78}{
\begin{tabular}{|c|c|c|c|c|c|}
\hline
\rotatebox[origin=c]{0}{Literature} &
\rotatebox[origin=c]{0}{Use Case} &
\rotatebox[origin=c]{0}{Storage Layer} &
\rotatebox[origin=c]{0}{Organisation Layer} &
\rotatebox[origin=c]{0}{Data Type Layer}
\\
\hline

 \cite{hyvarinen2017}& Finance &\checkmark	& \checkmark	& \checkmark	\\
 \hline
 \cite{andoni2019}& Energy &\checkmark	& \checkmark	& \checkmark	\\
 \hline
 \cite{choi2019}& Supply Chain &\checkmark	& \checkmark	& \checkmark	
 \\
 \hline
 \cite{leng2018}& Supply Chain &\checkmark	& \checkmark	& \checkmark	
 \\
 \hline
 \cite{perboli2018}& Supply Chain &\checkmark	& \checkmark	& \checkmark	\\
 \hline
 \cite{xu2019}& Supply Chain &\checkmark	& 	& 	
 \\
 \hline
 \cite{kuo2017}& Healthcare &\checkmark	& \checkmark 	& 	
 \\
 \hline
 \cite{Xu2020}& IoT &\checkmark	& 	& 	
 \\
 \hline
 \cite{mendki2019}& IoT &\checkmark	& 	& 
 \\
 \hline
 \cite{pustivsek2019}& IoT &\checkmark	& \checkmark	& \checkmark \\
 \hline
 \cite{guin2018}& IoT &\checkmark	& 	& \\
 \hline
 \cite{yang2018}& Vehicle networks &\checkmark	& 	& \\
 \hline
  \cite{samaniego2016}& IoT &\checkmark	& \checkmark  & \\
  \hline
  \cite{saberi2019}& Supply Chain &\checkmark	& \checkmark	& \checkmark \\
 \hline
  \cite{dorri2017}& IoT &\checkmark	& \checkmark	& \\
  \hline
 \cite{dai2017}& Finance &\checkmark    & \checkmark    & \checkmark \\
 \hline
 \cite{kshetri2018}& E-Voting &\checkmark	& \checkmark  & \\
 \hline
 \cite{zyskind2015}& Private Storage &\checkmark	& \checkmark  & \\
 \hline
  \cite{mendling2018}& Process Management &\checkmark	& \checkmark  & \checkmark \\
   \hline
    \cite{chen2019}& Healthcare &\checkmark	& \checkmark  & \checkmark \\
    \hline
    \cite{chen2020}& Energy &\checkmark	&   &  \\
\hline
\end{tabular}}
\break
\caption{Blockchain literature – The identification of which storage layers were mentioned in each implementation}
\label{tab:StorageLayers}
\end{center}
\end{table}

\textbf{\textit{Global supply chain solutions}} - In \cite{choi2019}, a blockchain is used to coordinate global supply chains, and manage risk, supply and demand relating to air logistics. In \cite{saberi2019}, a similar solution is presented where the logistics mode is generalised. The depth of the necessary blockchain implementation is lacking, however, the data type of smart contracts using the organisation of a distributed ledger is named. A blockchain is used for the integrity of orders through its immutability and the analytics of supply chain transactions. In \cite{leng2018}, an agricultural supply chain solution implements a dual blockchain public and private. The public blockchain serves the same purpose as in \cite{choi2019}, with the private blockchain being used to store confidential data. In \cite{perboli2018}, a blockchain for logistics and the supply chain uses improved communication, through data collaboration between parties, in a traditionally hierarchical layered relationship. In \cite{xu2019}, the supply chain is enhanced with a blockchain to establish the integrity of electronic hardware, and its supporting components and services. This design uses a distributed private blockchain, with an independent storage configuration within each block, that is, it is not in an organisation of a ledger, but segmented to allow for the verification of data.

\textbf{\textit{Health industry}} - In \cite{kuo2017}, focuses on the improved management of health records for both healthcare and biomedical applications, the authors seek to remove traditional centralised database management systems (DBMS) – the software that organises data into information through the use of a query language. In this solution, the desired aspects of blockchain are its lack of centralised governance for record access and the ability to audit chronologically and provide immutability for information stored. Also, this solution highlights the availability and robustness of data stored as distributed replica copies. It can be inferred that the data is stored via a transaction within an organisation structure of a ledger. That is each update to a medical record is a stored with other entries within a block, this being its data type as presented in figure  \ref{fig:blockchainstoragecomponents}. The complete medical record is to span multiple blocks as the sum of all the entries within the blockchain. The purpose of a blockchain in this solution is the availability of health record information. In \cite{chen2019}, a blockchain is used to store an index of information. This includes a scheme for accessing encrypted health records stored externally to the blockchain across independent data storage platforms. A smart contract data structure is used to manage the financial aspects of the system, such as fees for record access.

\textbf{\textit{Internet of Things (IoT)}} - 
In \cite{Xu2020} and \cite{mendki2019}, the authors used blockchain as an extension to the IoT and mobile edge devices as a storage solution with the focus on removing the storage overheads from traditionally low powered devices. This forms a replicated blockchain state over a distributed network topology, which incorporates the compute overheads required to maintain the necessary read-write latency, integrity, security and availability for the IoT devices’ data storage. Similarly, in \cite{pustivsek2019}, the authors demonstrate a solution to reduce overheads required by IoT devices, especially using the Ethereum blockchain architecture, which infers the use of the data type smart contracts. In \cite{samaniego2016}, a blockchain performance comparison is carried out between a cloud platform blockchain and a private blockchain to determine acceptable latency for IoT devices. In this solution, a distributed ledger organisation structure is implemented, however, core blockchain specifics are lacking. In \cite{guin2018}, a blockchain is implemented for authorisation and authentication of IoT devices in a two-step process, first a public blockchain holds the registered manufacture ID for an IoT device. In the second step, a private blockchain, with authorised mining nodes, manages the registration of each device. Both configurations use a distributed blockchain to store a unique transaction structure for its data type, related to its role in authorisation and authentication of an IoT device within the network. Furthermore, in \cite{dorri2017}, a similar blockchain solution is implemented for IoT devices, which includes detail about the use of core blockchain technologies.

\textbf{\textit{Vehicle networks}} - In \cite{yang2018}, the solution uses two types of blockchains – one is for a restricted group’s communication and the other for broadcast messages, where selected vehicles act as mining nodes. These nodes add new transactions obtained from vehicles within its range. The use of blockchain in this example is as storage, which accepts a file as the data type this being presented in figure \ref{fig:blockchainstoragecomponents} as a file object. This solution makes use of the distributed storage replication for improved availability.

\textbf{\textit{E-voting}} - In \cite{kshetri2018}, blockchain in e-voting uses a distributed ledger organisation structure presented in figure  \ref{fig:blockchainstoragecomponents}, with tokens or coins being issued and later used to cast votes. Technical detail is lacking about the fundamental breakdown of the technologies used.

\textbf{\textit{Personal data storage}} - In \cite{zyskind2015}, the blockchain’s purpose is to store personal data or ID data in an encrypted state. The blockchain uses a custom transaction data type for storage within a distributed ledger organisation structure. The technologies used in the blockchain are discussed, including the external supporting systems that allow for data exchange.

\textbf{\textit{Business process management}} - In \cite{mendling2018}, a review is done directly relating to the implementation of blockchain and smart contract data structures, as presented in figure  \ref{fig:blockchainstoragecomponents}, for the execution of business processes. The general technical details of blockchain are explored, but the specific implementation is not discussed.

What can be concluded from the related works reviewed is that each paper refers to blockchain as a solution to a real world challenge, however, the identification and classification of blockchain is often lacking. Moreover, this is highlighted by the difference in the solution characteristics. Table \ref{tab:StorageLayers} identifies which storage layers (based on figure \ref{fig:blockchainstoragecomponents}) are discussed in each blockchain paper. This highlights that the term blockchain always includes components of the storage layer, with components at the organisation layer and data type layer being associated and optional technologies for inclusion. Moreover, blockchain can be defined and identified by the technologies that make it an immutable storage solution, only within the storage layer, shown in figure \ref{fig:blockchainstoragecomponents}.

\begin{figure}[t]
    \begin{center}
    \includegraphics[width=0.75 \columnwidth]{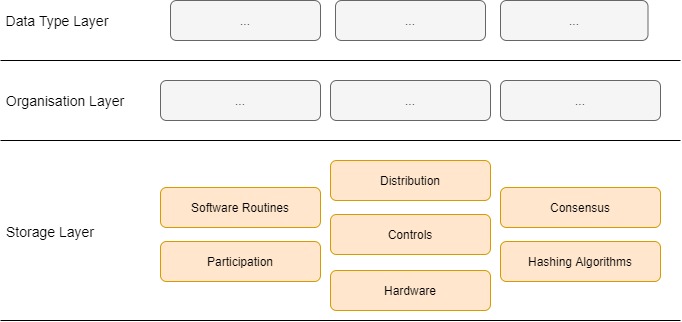}
    \caption{{Blockchain components in the storage layer.}}
    \label{fig:blockchaincomponentslayers}
    \end{center}
\end{figure}

\section{Blockchain Components}
\label{blockchaincomponetsdefined}

To identify the core blockchain components we introduce an extension to the storage layer model presented in figure \ref{fig:blockchainstoragecomponents}, where we move from the design construction of a blockchain to a logical structure in figure \ref{fig:blockchaincomponentslayers}. This structure consist of seven groups of components that make up a blockchain existing solely at the storage layer. Every blockchain can be defined by which components are implemented, moreover, these groups generalise the necessary components to construct a blockchain. As current blockchain solutions remain application specific, each group contains the components, technologies or architectural considerations that exist as either a working implementation or within research articles. In figure \ref{fig:blockchaincomponentslayers} is the storage layer components, any layers above this we consider a blockchain’s associated components, technologies or architectural considerations, expressed within the organisation layer and the data type layer, presented in figure \ref{fig:blockchainstoragecomponents}. In the following sections we explore each of the seven component groups at the storage layer of our novel blockchain component model.

\subsection{The Storage Layer}
\label{blockchainstoragelayer}

The storage layer, in figure \ref{fig:blockchaincomponentslayers} presents the components that we consider a blockchain's core components - a generalised representation of components found in blockchain implementations. The seven components presented cover the topics of the blockchain software, the privacy and a varying level of security, discussed in the participation types in a blockchain, and the distribution of storage through the replication of the blockchain. We continue with the components of  the controls which manage the blockchain's function, the consensus method that coordinates the blockchain's participation, the hashing algorithms that give immutability and functionality to the blockchain, and finally the specific blockchain computer hardware. Each of these components is explored in the following sections.

\subsubsection{Software Routines}
\label{softwareroutines}

We begin with the software routines component within the storage layer of our three-tiered blockchain model, presented in figure \ref{fig:blockchaincomponentslayers}. Blockchain typically can be considered a replicated state machine for recording data in an immutable manner, this is achieved through the use of software installed on each of the participants computers. To control, operate, and coordinate this replicated state machine, a collection of algorithms, procedures and technologies layout the function and design of the blockchain \cite{xu2016}. A programming language needs to be selected that supports the core components required in the blockchain design, such as the components for the Bitcoin blockchain, which were discussed in section \ref{thebeginning}. As an application suite, the blockchain software routines, can be broken into server or node components and client components used to interact with the blockchain and its supporting network. It is this selection of software with which the blockchain is started, utilised and maintained during its operation.

Not unique to blockchain is the software development lifecycle (SDLC); this brings about new challenges regarding centralised or distributed control, coordination of resources, communication and interaction of blockchain stakeholders, who all contribute to the successful operation of the blockchain \cite{ruparelia2010}. This is especially apparent when a challenge arises with a functioning blockchain or its network. An example of this occurred in 2017 when the Bitcoin blockchain suffered from design constraints and excessive transaction processing times. This resulted in the emergence of two solutions, the Bitcoin Core blockchain and the Bitcoin Cash blockchain. A difference in the blockchain software development direction caused a schism between the supporting decentralised developers \cite{kwon2019}. On the contrary, a private blockchain with centralised developers is unlikely to have a similar result. This is especially true when a centralisation of control and authority is desirable in a blockchain solution. Blockchain software routines define a blockchain as storage and the supporting network operations – they are therefore considered core blockchain components, highlighted in figure \ref{fig:blockchaincomponentslayers}.

\subsubsection{Participation}
\label{participation}

Referencing our three-tiered blockchain model we next consider the participation component in the storage layer, presented in figure \ref{fig:blockchaincomponentslayers}. Blockchain participation can be divided into two main areas, the first is client participation, that is, who has access to the data on the blockchain. Access being read access to existing data and write access to queue data for inclusion in the next block \cite{zheng2017}. The second area is the server or node access, that is, what authentication is required to participate in the blockchain network for block addition onto the blockchain and the supporting tasks such as transaction replication. These two roles can be identified in figure \ref{fig:blocktransaction}, with the clients being the two computers on the left of the figure conducting a transaction and the server or node being part of a mining pool or separate mining nodes. However, the role or task is not necessarily limited to mining.

Blockchain participation requirements are defined as either being a public blockchain, having no authentication requirements for participation, or a private blockchain, having client and server participation requiring authentication. A blockchain can also consist of a variation of authentication required for the client’s and the server’s participation. This can be for both the data stored on the blockchain and the functions of the blockchain network \cite{dinh2018}. This is sometimes referred to as a hybrid model and when authentication is required, this is referred to as a permissioned blockchain. The type of blockchain generally has a relationship with the consensus methods used, described in section \ref{consensusmethods}, as some consensus types are not desirable for the participation types previously stated, due to the tradeoff between security and the rate at which blocks can be added \cite{Gupta2020}. The participation within the blockchain network and access to the data on the blockchain are considered core components, shown in figure \ref{fig:blockchaincomponentslayers} and a necessary consideration in a blockchain’s architecture.

\subsubsection{Distribution}
\label{distribution}

The next component of our storage layer model, presented in \ref{fig:blockchaincomponentslayers}, is the distribution and replication of the blockchain throughout its network, and the computer storage requirement for the blockchain. The blockchain storage distribution is a consideration that is determined by elements of its participation in section \ref{participation}. Peer-to-peer computer storage replication has been a popular blockchain storage replication protocol for use with public blockchain participation over the Internet. This can be broken into two designs; first, unstructured distribution using methods such as Gossip protocols, where it is assumed that all known neighbors have a uniform latency or distance \cite{shahsavari2019}. The second distribution design is structured, with each neighbor’s network properties, such as latency or distance, being calculated and used as a network map \cite{sallal2017}. This network topology creates efficiencies in consensus, network usage, replication rate, transaction addition and transaction verification rate.
 
In addition to the network topology is the storage requirements of the blockchain. Some blockchain configurations allow for different purpose nodes to reduce the storage overheads within the network. This is achieved by requiring only some of the nodes to retain complete copies of the blockchain. Private blockchains can benefit by using traditional shared storage technologies, such as a storage area network utilising low latency connectivity, such as a fiber channel and low latency local area network \cite{Sukhwani2017}. This can eliminate some of the undesirable temporal traits through the centralisation of a blockchain.

As a universal consideration, distribution requires a set of protocols to engage in tasks for the acceptance of new blocks, replicating existing blocks, node name resolution, and replicating transactions between nodes. These protocols allow for nodes to serve blockchain information and receive blockchain information \cite{sudhan2018}. This is defined in the blockchain’s software routines in section
\ref{softwareroutines} and is executed based on the participation model explored in section \ref{participation}. This is considered another core component in a blockchain’s design, displayed in figure \ref{fig:blockchaincomponentslayers}.

\subsubsection{Controls}
\label{controls}
\begin{figure}[t]
    \begin{center}
    \includegraphics[width=0.75 \columnwidth]{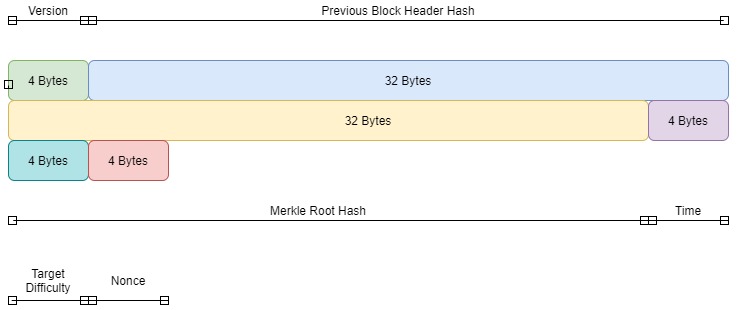}
    \caption{{The Bitcoin blockchain block header layout.}}
    \label{fig:blockheaderlayout}
    \end{center}
\end{figure}

From our three-tiered model we next consider the component in the storage layer called controls, presented in figure \ref{fig:blockchaincomponentslayers}, these are used to manage aspects of a blockchain's workings. At the time of writing, controls used in a blockchain remained application specific. One common control used in blockchains is the implementation of temporal limitations \cite{fullmer2018}. This type of control is often implemented within the blockchain’s consensus algorithm discussed in detail in section \ref{consensusmethods}. The consensus algorithm PoW has control parameters adjusted within the header of each block, expressed as the target difficulty in figure \ref{fig:blockheaderlayout}. This control functions by adjusting a difficulty value for the complexity of the proof required, in doing so the blockchain network maintains an average time of ten minutes between block additions, expressed in the upper portion of figure \ref{fig:blockchainminingmechanics}. The purpose of this is for currency control; by enforcing a limit to the growth rate of the blockchain, this in turn limits the amount of Bitcoin cryptocurrency released in a given period. Another example of a blockchain control, expressed as a temporal limitation, is during the operation of a blockchain consensus algorithm called Proof-of-Elapsed Time (PoET), a temporal control is enforced by specialised hardware. This is designed to invoke a wait in block processing for a defined period \cite{chen2017}.

Another example of an application specific blockchain control, which is not a temporal limitation, is in the Bitcoin blockchain. This is the halving of the amount of Bitcoin cryptocurrency released every cycle; this cycle being the addition of every 210 thousand blocks. From first set amount being defined in the Genesis block of the blockchain, presented in the top section of figure \ref{fig:blockchainminingmechanics}, this cycle has continued to enforce this control. The control is implemented by the parameter of the Genesis block, and in the software routines, explored in section \ref{softwareroutines}, and is expressed in each block of the blockchain.

The controls mentioned above are application specific and are undesirable for some blockchain implementations, such as collecting votes for a vast population in a limited period, in \cite{khan2020}, or the registration of health information during an active pandemic. Blockchain controls are unlikely to be generalised, however, they are a consideration when designing a blockchain especially regarding the design of the organisation layer and data type layer, highlighted in figure \ref{fig:blockchainstoragecomponents}. Blockchain controls exist within the storage layer and are considered a core blockchain component, presented in figure \ref{fig:blockchaincomponentslayers}.

\subsubsection{Consensus Methods}
\label{consensusmethods}

From our blockchain layer model we discuss the next component in the storage layer called consensus, presented in figure \ref{fig:blockchaincomponentslayers}. The purpose of a consensus method in a network of nodes is to establish a protocol for the negotiation of how each node will function to achieve a single task. This is regardless of the usefulness or willingness of each node to do so \cite{barborak1993}. This is commonly referred to as a Byzantine agreement, which assumes the existence of malicious or intermittent nodes. Consensus forms coordination of the network nodes.

In a blockchain network, the consensus method is explicitly used to agree on what blocks are added to a blockchain, and is defined in the building of the software routines, described in section \ref{softwareroutines}. When viewed at a high level, the consensus methods used are selected based on the characteristic of what blockchain network participation type has been selected, explored in section \ref{participation} and the blockchain method for distribution selected explored in section \ref{distribution}. A general criteria for blockchain consensus selection is the ability for the network to grow or scale, such scalability constraints are inherent in some consensus methods \cite{xiao2020}.

Blockchain consensus methods are numerous covered in detail in  \cite{chaudhry2018} and \cite{xiao2020}. An identification, explanation and examination of current consensus methods in relation to blockchain have been extensively documented. A general overview is that public blockchain consensus methods, such as PoW, trade speed for security to function between unverified and unauthenticated network nodes. Therefore consensus convergence is expected to be delayed. On the contrary, a consensus method such as a variant of a Byzantine Fault Tolerant (BFT) algorithm is often used to negotiate orderly block addition and replication, and trades security for speed. This is often used for nodes in a private blockchain network, which are authorised and authenticated before their participation and therefore the consensus convergence is rapid \cite{gramoli2020} \cite{kotla2004}. In must be noted that Byzantine Fault Tolerance like methods for consensus are not unique to blockchain and have been in existence prior to blockchain \cite{Castro2002}. In the following section will seek a different approach to investigate some commonly available open frameworks, projects and source code used for blockchain development, where a traditional block construction is implemented. We will identify each consensus method and how it functions.

\begin{figure*}[t]
    \begin{center}
    \includegraphics[width=0.9 \columnwidth]{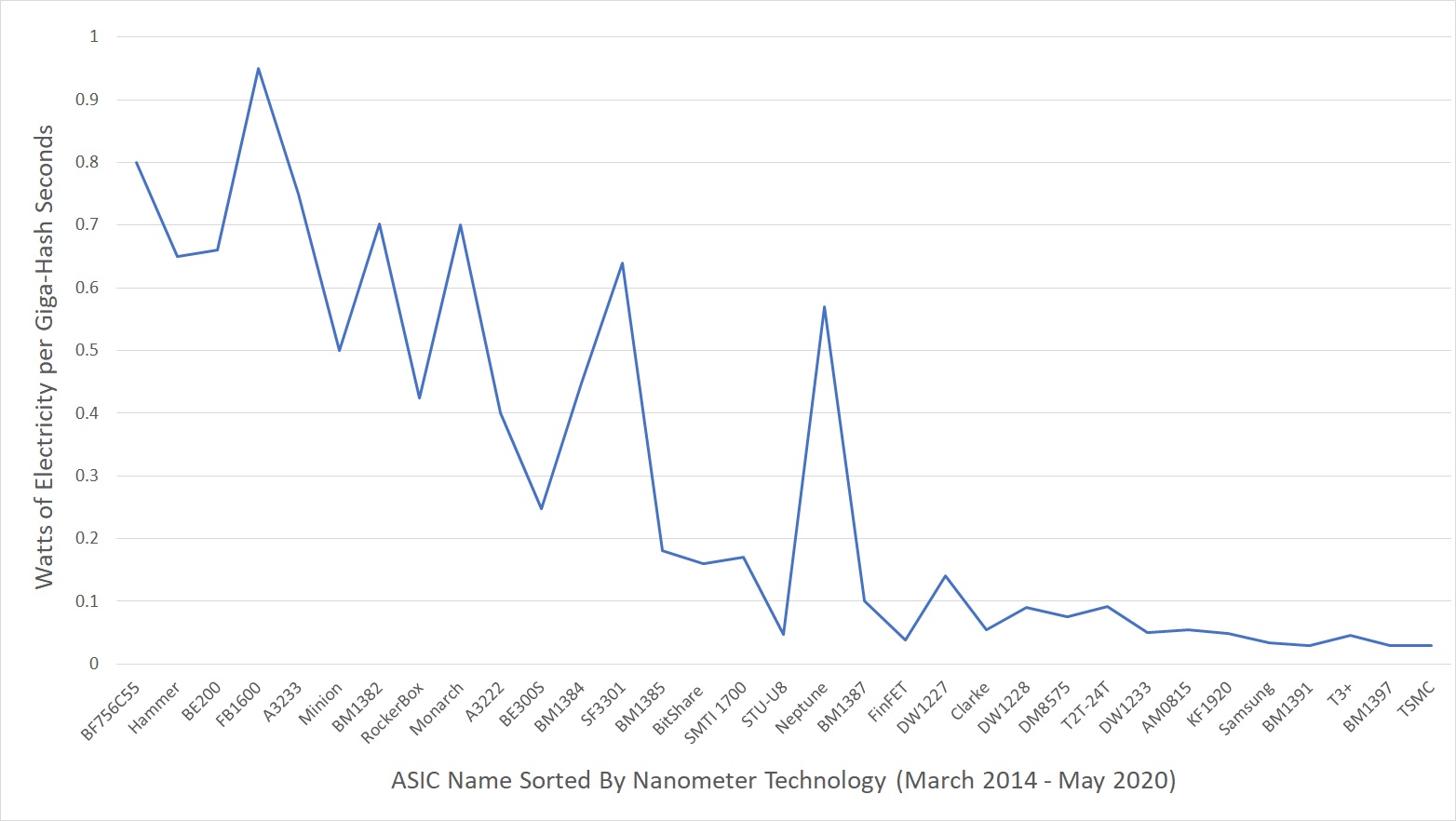}
    \caption{{ASIC evolution for blockchains using SHA-256 hash function from the year 2014 until 2020. Presenting the watts of electricity required for 1000 hash calculations every second. Ordered from left to right by the nanometer technology used.}}
    \label{fig:asicevolution}
    \end{center}
\end{figure*}

Open Frameworks and Projects – First is the Bitcoin core blockchain project review, in \cite{bitcoin_core}, which uses a PoW consensus method. A significantly more compute intensive PoW consensus method is used; the work being a solution to a mathematical puzzle \cite{Lewenberg2015}. The consensus method consists of the production of the proof or block seal. The requirement for the proof in the Bitcoin core blockchain, in January 2021, resulted in the network producing a little over $166\times10^{18}$ attempts to solve the puzzle each second. By using a spread of ASIC technology from this period, presented in figure \ref{fig:asicevolution}, it can be estimated that 4.7 and 26 gigawatts of electricity was required every second to maintain the blockchain. This energy usage is directly related to the total capacity of the mining nodes in the network, and therefore this consensus method is considered energy intensive. However, this is not constant throughout the operation of the blockchain and is a result of the controls discussed in section \ref{controls}. 

For the consensus method of PoW on Bitcoin Core blockchain we look into how this is implemented. The values which are used to determine a successful block proof are stored in each block’s header value, displayed in figure \ref{fig:blockheaderlayout}, and as parameters in the 'coinbase' transaction. The first transaction of a block which is created by the mining nodes governed by  software routines discussed in \ref{softwareroutines}. Each block is tested by all the nodes receiving the block update by a calculation of the blocks parameters. An example of an accepted proof is at the block numbered 632,462, with the truncated proof or hash starting with ‘0000000000000000000a15...’. The number of leading zeros is the result of the required difficulty expressed in the 4 byte location of the header. This hash was found using the nonce value of 2,459,281,113. The nounce value is expressed in both a 4 byte location in the header and in the extraNonce parameter of the coinbase transaction. The resulting proof or hash is produced from all the 6 areas of the block header. 

PoW is also probabilistic – it is possible that more than one node can locate a suitable hash, having an equal or greater number of leading zeros. When this occurs, the nature of the Gossip communication protocol used results in an imbalance of the updates announcements throughout the network. This results in one version of the blockchain being longer than the other. The longest version of the chain takes precedence, which is the version held by greater than 50 percent of the network. The shorter version is labelled a fork in the blockchain and is ignored. As a result, the block and its transactions are discarded. Due to this, transaction confirmation is not instant and works on the principle that the more blocks that are added, the less likely that the block and the transaction will be discarded \cite{Decker2016}. The implementation of PoW consensus using this method allows for a globally sparse network of an ad-hoc number of participants.

The second is the Hyperledger project, which consists of a modular approach to a blockchain implementation \cite{Hyperledger}. We will look into two solutions within this project, Indy and Sawtooth and the consensus method applied in each. Indy is a private blockchain solution that implements a Redundant Byzantine Fault Tolerance (RBFT) based protocol to reach consensus \cite{aublin2013}. It is acknowledged that a private blockchain configuration is implemented in combination with a centralised method for access control, and therefore, the trust in a node is established before its participation. Due to this, a single node is only required to be selected to create a seal for the block. This results in a minimal amount of computational power to maintain the blockchain, and allows for accelerated block addition. In RBFT, an implementation of Byzantine Fault Tolerance is used to select a primary node for processing the block. This process is replicated to produce a backup primary node, which is a node that hasn’t been selected as a primary or backup primary. RBFT can include many backup primaries, which perform monitoring of the primary to determine if it has failed. All messages are signed and processed by all nodes, but applied to the blockchain only by the primary \cite{Indy-plenum}. The backup nodes compare speed and accuracy to establish trust in the primary. The primary node will seal the block using the Boneh–Lynn–Shacham signature scheme \cite{boneh2004}, resulting in a 170-bit seal of all the nodes involved in consensus.

Sawtooth uses a different consensus method, called Proof-of-Elapse-Time (PoET) consensus \cite{olson2018}. PoET trust is established through the secure execution of the algorithm within a proprietary environment, using specialised hardware. Therefore, trust is centralised around the manufacture of the hardware. PoET requires that each hardware node registers using a key pair, thus forming a permission blockchain. PoET works by each node in the network calculating a wait time, which follows a statistical distribution for minimising two devices’ wait times colliding. A wait time calculation can be used a set number of times before it has to be recalculated, counting backwards from 25 to 0. Each node waits for its specific time before adding a block to the blockchain. All nodes do verification in the network before the block is accepted. The blockchain is statistically analysed to reveal any anomalies with the participating nodes \cite{chen2017}.

The third is Ethereum 2.0 and the Beacon Chain, which implements a Proof-of-Stake (PoS) consensus model \cite{Ethereum_POS}. This method considers that an entity that has a significant ownership of the cryptocurrency is trusted to participate in the addition of blocks to the blockchain. This stake is forfeited or discounted in the event of malicious participation. From the group of validator nodes, each one providing its stake, one is selected via an algorithm to carry out the block addition. Each block is signed by the selected node using a Boneh–Lynn–Shacham signature scheme. Verification of blocks added is then carried out by the validator group. This group consists of 128 members called a committee and works on their associated block in a sub-unit called a shard \cite{Zamani2018}. This model results in significantly less computational resources to operate, and removes the incentive for centralised mining pools within the network, in \cite{MaungMaungThin2018}.

The fourth is the Graphene blockchain, which uses a Delegated Proof-of-Stake (DPoS) consensus model \cite{Graphene}. All aspects of a proof-of-stake consensus are applied. However, the difference is the removal of the algorithm for the selection of the validator node. This is replaced with the nodes voting for a node which is tasked with the addition of the next block. The consensus modifications improve the speed at which blocks can be processed \cite{Zhang2020}.

The fifth is Ethereum’s mainnet blockchain, which uses a memory intensive PoW consensus method \cite{Ethereum_POW}. This differs from the Bitcoin blockchain consensus in several ways. The transactions are stored in a ledger, with each block having a complete state of ledger stored as a Merkel tree, displayed in figure \ref{fig:binarymerkletree}. Ethereum’s mainnet uses a hashing algorithm called Keccak-256, which is a modified implementation of SHA3-256. Hashes are performed on a random selection of state and transaction data in the blockchain – because of this, the whole blockchain needs to be available in memory. The PoW is finding a nonce value, which produces a hash with the required difficulty specified in the block header. This approach assists to limit the usefulness of ASIC chips in the PoW mining process, due to the insufficient memory included in ASIC chips. Once the block is complete, it is announced to other mining nodes and verified. The longest blockchain is maintained, with any forks being discarded, in the event a block solution is found by more than one node \cite{gervais2016}.

The last is the Exonum blockchain, which uses a Practical Byzantine Fault Tolerance (PBFT) with some customisation for its consensus model \cite{Bitfury_Exonum}. This requires the network of nodes to be partially synchronous, as the consensus steps are processed in rounds and coordinated through the use of application messages. When a signed message arrives, it is verified and its transaction is added to a transaction queue. Voting is carried out and accepted if two-thirds or more of the nodes accept the transactions to be included in the block and accept the node for processing the block. This selected node is given the title leader and issued a Proof-of-Lock (PoL), required for block addition, which serves a similar purpose to a transaction lock in a database. Once the items have been added to the block, they are then deleted from the transaction queue. These steps are repeated to form the next block addition cycle.
 
 \textit{Generalised Consensus Methods} – In the following, other solutions to consensus within a blockchain network are reviewed. A number of PoS variations exist, such as Leased Proof-of-Stake (LPoS), in which trust can be leased from an owner of a cryptocurrency, and Proof-of-Stake Anonymous (PoSA), which has a focus on obfuscating a transaction’s inputs and outputs. Proof-of-Stake Time (PoST) is based on how long the cryptocurrency has remained at an account or wallet address. The consensus model Proof-of-Burn (PoB), in \cite{karantias2020}, establishes trust through the owner of the node disposing of cryptocurrency. This works as surety that the node is honest and that the block and transactions are formed correctly. The consensus model Proof-of-Importance (PoI) \cite{Sankar2017}, takes into account the amount of cryptocurrency, how many transactions are carried out using the associated wallet or account address, and the system’s configuration. The consensus model Proof-of-Luck (PoL) \cite{Milutinovic2016}, makes use of a trusted execution environment or specialised hardware. A chain is processed in rounds, where communication on the luck factor calculation is taken into account. Certain values are considered desirable, which result in establishing a preferred block to be included in the blockchain, other blocks are discarded. Proof-of-human-work requires that the solution to a computationally verifiable problem is done by a person \cite{Blocki2016}. Proof-of-useful-work returns a prime number value, which can be used in applications outside of the blockchain consensus method \cite{Loe2018}.
 
 The threshold relay model consensus method \cite{dfinity_consensus}, works by using a verifiable random function to produce a signing token, which is then broadcast to the network. Each node in the network with its designated role produces a block using this token. Once the node creates the block, it is returned to another decentralised process, labelled the notoriety, which receives all the blocks. A block is selected and broadcast as the next successful block in the chain through a set of rules. It has to be noted that the consensus methods explored above do not encompass every known type, but seek to highlight some differences in blockchain consensus implementations. Consensus methods are considered core components of a blockchain and are located at the storage layer, in figure \ref{fig:blockchaincomponentslayers}.
 
 \subsubsection{Hashing Algorithms}
\label{blockchainconsensushashingalgorithm}

The next component of our blockchain layer model is the hashing algorithms exiting in the storage layer, presented in figure \ref{fig:blockchaincomponentslayers}. Hashing algorithms have the primary function to form cohesion of the blocks, or linking in a blockchain. In this section we explore the implementations and hashing algorithms used in blockchain solutions. A hashing algorithm is a computer implementation of a hash function and can be generalised to represent data in a unique encoding scheme \cite{chi2017}. Blockchains make use of cryptographic hashing, where the encoding of an arbitrary amount of data called a message, not exceeding the function’s upper bounds, results in a fixed size code called the message digest. It is considered not possible to reverse this process to expose the original message \cite{wang2018}. Depending on the blockchain’s implementation, the selected purpose of a hash function can be broken into the fundamental requirements for a blockchain. First, a hashing function needs to be implemented to ensure data contained within a block remains unchanged, by verifying its hash representation. Second, when a puzzle is used, generally in PoW, it provides consensus when joining blocks together, explored in section \ref{consensusmethods}. In both cases, the hash function selected needs to be collision resistant, that is, only one message is represented by exactly one message-digest. In addition, different hash functions can be used in one blockchain implementation. Furthermore, hashing can also be used on the data stored within each block, which exist at the data type layer in figure \ref{fig:blockchainstoragecomponents}, such as in a wallet’s address, which is a message digest of the public key, \cite{dasgupta2019}.

Blockchains, being application specific, have selected their hash functions based on other criteria. One of these criterion is to remove the participants using ASIC mining, by implementing multiple hash functions in the mining process. The output of one hash function is passed to the input of another hash function, in a chain of different hash functions. X11, X12, X13, X14, X15, X16R, X16S, X17 follow this to include a selection of hashes from Blake, Blue Midnight With, Cubehash, Djb2, ECHO, Fugue, Grøstl, Hamsi, JH, Keccak, Loselose, Luffa, SHA2-512, Shabal, Shavite-3, SIMD, Skein, and Whirlpool. Other implementations have used a similar approach to achieve ASIC mining resistance, with the preferred method of CPU or GPU mining \cite{Cho2019}. The reason for this exclusion is to decentralise the mining processes by reducing the cost of hardware outlay for participants, with varying levels of success. Also, it allows more participation and therefore the more likely adoption of the blockchain and its application. Another criterion is the hash function’s performance, such as comparing of the SHA2 family of hash functions 256, 384, and 512 \cite{Sklavos2003}, with blockchain implementations maintaining a preference for SHA2-256. A comparison of blockchain’s hash function concerning performance is explored in \cite{kuznetsov2019}. The amount of data to be hashed is a consideration for both the message size limits of the hash function and the estimated time to calculate a message digest. 

Hash functions used in a blockchain are used nearly universally for data validation. However, some block seals are implemented using a different technique of signing, which is explored in section \ref{consensusmethods}. Hashing components are core to a blockchain solution and therefore exist at the storage layer, in figure \ref{fig:blockchaincomponentslayers}.

\subsubsection{Hardware}
\label{hardware}
From our three-tiered layer model we present the last component in the storage layer labelled hardware, presented in figure \ref{fig:blockchaincomponentslayers}. Blockchain specific hardware development has been historically driven by the desire to accelerate the rate at which a successful hash can be located \cite{Vilim2016}. However, it must be noted that not all consensus algorithms and therefore blockchains require specialised hardware \cite{han2019}. This is due to being intensive in node memory and not in node compute cycles. A non-blockchain specific GPU is used to generate the proof in some implementations. Some blockchain consensus methods make it advantageous to incorporate parallel processing using application specific integrated circuits (ASIC), explored in section \ref{consensusmethods}. ASIC are specially designed hardware used in the blockchain block addition process, also known as mining. These ASIC are designed around a particular hashing function, previously discussed in section \ref{blockchainconsensushashingalgorithm}. Application specific blockchain hardware has been generally driven by the incentive of financial reward of the respective blockchain cryptocurrency. This is best explained using Bitcoin’s blockchain, specially the PoW consensus using SHA-256 hash function, and its evolution.

The Bitcoin blockchain's PoW consensus method has resulted in four main generations of blockchain mining solutions. As this technology was novel, block calculations and the search for a successful hash was done with software. This first generation of blockchain mining node used the computer’s central processing unit (CPU). This was the most cost-effective way to process blocks relative to the financial reward. As the interest in the technology spread, so too did innovation of faster ways to calculate a successful block hash and the likelihood of receiving the associated financial reward. This resulted in the second generation of mining, which made use of graphic processing units (GPU). This resulted in significant improvements to processing hashes \cite{kuznetsov2019}. As these methods became too ineffective and inefficient in terms of the rate of processing hashes per second, a move to specifically configured hardware resulted; this being the third generation of mining using field-programmable gate arrays (FPGA) \cite{SOLIMAN2011}.

FPGAs were used as a decoupled hardware solution. These devices allow the hardware to be programmed to perform a specific task, such as the routine required for blockchain mining, and are specially designed for the SHA-256 hash function \cite{devika2019}. The difficulty parameter continued to grow and made this solution obsolete. This resulted in the fourth and current generation of mining hardware identified by the use of application specific integrated circuits (ASIC). These chips are designed with the algorithm implemented and fixed at the chip  manufacturing stage. 

ASIC technology has continued to be developed with improvements in power efficiency and an increase in the rate at which hashes are calculated. In table \ref{tab:HashingChips}, the type of hash function deployed as an ASIC chip is tabulated. The column ‘maintained’ identifies if the ASIC for the hash function is still under development. As a continuation of the work in \cite{taylor2017}, it can be observed that the opportunities have diminished to implement novel ideas for executing the algorithm, such as it was during the first to third generations of mining and in the early years of the fourth generation. Figure \ref{fig:asicevolution} shows the number of watts of electricity for 1000 hash calculations per second, sorted by the ASIC technology evolution from 55 nanometers until the current designs of 7 nanometers, between March 2014 and May 2020. This highlights that comparatively smaller gains are achieved over time, furthermore, the gains today are directly associated with which ASIC nanometer technology is used to build the chip. Such ASIC's are then incorporated into hardware units consisting of hundreds ASIC chips, often called mining rigs.

Specific blockchain hardware has also included propriety hardware, specifically designed for certain consensus methods such as PoET discussed in section \ref{consensusmethods}. In addition, some implementations continue to use generalised CPU and GPU hardware in the block addition process. It is acknowledge that hardware is critical for all blockchain solutions and we consider it a core component existing at the storage layer, in figure \ref{fig:blockchaincomponentslayers}.

\subsection{Organisation Layer}
\label{blockchainorganisationlayer}
The organisation layer and the components defined are considered as associated blockchain technology, presented in our model in figure \ref{fig:blockchainstoragecomponents}. Moreover components of this layer are considered optional to use when building an immutable storage solution, as highlighted by the omission of the organisational layer in some blockchain implementations, presented in table \ref{tab:StorageLayers}. Furthermore, data can be stored on a blockchain without any organisation. This is due to the blockchain properties at the storage layer; such as chronological ordering or immutability of data, which may be sufficient for a solution. Moreover, the organisation of data may be performed by the software application using the blockchain for its storage. However, this layer is often used to structure the block data and forms consistency across all blocks in the blockchain, with the distributed ledger being the most commonly used data structure to date. In practice, there is no limitation to what data structure can be implemented. In this section we will explore the types of data structures existing at the organisation layer.


\begin{table}[t]
\centering
\begin{tabular}{|l|l|l|l|}
\hline
Hash Function Name & Proof Type & ASIC Implementation & Maintained               \\ \hline
SHA-256            & PoW        &      \checkmark               & \checkmark \\ \hline
X11                & PoW        &        \checkmark             & \checkmark \\ \hline
Ethash SHA3-512    & PoW        &         \checkmark            &  \\ \hline
Equihash           & PoW        &       \checkmark              & \checkmark  \\ \hline
CryptoNight        & PoW        &       \checkmark              &                          \\ \hline
Scrypt             & PoW        &        \checkmark             &                          \\ \hline
\end{tabular}
\break
\caption{Implementation as an ASIC chip.}
\label{tab:HashingChips}
\end{table}

\subsubsection{Ledger Structure}
\label{ledgerstructure}

We begin with the organisation layer type of the ledger, presented in figure \ref{fig:blockchainstoragecomponents}. Blockchain in its original application of Bitcoin implemented a ledger to manage the cryptocurrency in existence. The ledger is updated using transactions, including the change that will occur. In the case of Bitcoin, it is the change of ownership of the cryptocurrency, this  process is presented on the left of figure \ref{fig:blocktransaction}. This transaction is then recorded in the ledger through the blockchain mining process presented in figure \ref{fig:blockchainminingmechanics}. Furthermore, ledgers are not limited to financial transactions and can extend to any update in the form of a transaction. The term Distributed Ledger Technology (DLT) has also been used to describe a blockchain, with an organisation structure of a ledger. However, not all distributed ledger technology makes use of a chaining of blocks; some implement a chaining of transactions to form a ledger \cite{Kannengie2020}. Overall distributed ledger technologies focus on immutability and the rate at which transactions can be processed in a centralised environment, in comparison to blockchain ledgers, which are generally security focused and commonly designed for a decentralised environment, previously discussed in the section \ref{participation}. Ledgers are optional components that organise data contained within each block to produce an association of data across all blocks, and are placed in the organisation layer of figure \ref{fig:blockchainstoragecomponents}.

\subsubsection{Database Structure}
\label{datastructure}

Blockchain has been implemented to make use of relational and NoSQL database management systems, to improve information access when storing data on a blockchain. Requests are processed through a database management system. However, the database is stored as a blockchain. Therefore, maintains the properties of a blockchain, such as immutability, replication, and security in a decentralised environment \cite{Senthil2019}. Blockchain in itself can have its data organised to reflect a database structure natively, supporting traditional functions of database management systems, such as query language support and indexing, thus removing the need for a database management system \cite{zhu2019}. This type of structure organises data both in and across the blocks and therefore exists at the organisation layer, in figure \ref{fig:blockchainstoragecomponents}.

\subsubsection{Other Structures}
\label{otherstructure}

Blockchain implementations have included structures for storing and updating medical records, for IoT data storage, which varies on the type of device sending the data, to support structures for the authentication and authorisation of devices, and for traditional raw data organisation, in section \ref{relatedworkusingblockchainasstorage}. Technical specification of such solutions often lacks the details of exactly what structure it would require at the organisation layer, and proves to be application specific. However, this section is to acknowledge the existence of organisation structures, which have been implemented or purposed but are not shared or defined. As such structures are for the organisation of data between blocks, they are placed in the organisation layer, in figure \ref{fig:blockchainstoragecomponents}.

\subsection{Data Type Layer}
\label{blockchaindatatypelayer}

The data type layer, in figure \ref{fig:blockchainstoragecomponents}, is used to establish the purpose of the blockchain by defining what data will be stored. This layer establishes the application specific use of the blockchain, identified by what type of information and in what format it is to be stored. This is akin to file types in a file system such as MP3, GIF and PEM. Components of this layer are considered optional to use when building an immutable storage solution, as highlighted by the omission of the data type layer in some blockchain implementations \cite{dinh2018}, and presented in table \ref{tab:StorageLayers}. It is acknowledged that a blockchain primary purpose is for data storage, however, we consider that the structure of data being stored is not part of the core components required to produce an immutable storage solution, that is blockchain, presented in figure \ref{fig:blockchaincomponentslayers}. Moreover, a blockchain will continue to process blocks even if the data section of the block is empty. We define what types of data are commonly implemented at the data type layer of our model in the following sections.

\subsubsection{Smart-Contracts}
\label{smartcontracts}

At the data type layer of our three-tiered model, in figure \ref{fig:blockchainstoragecomponents}, we begin with smart contracts. Smart contracts are applications stored on a blockchain and are executed externally to the blockchain on a supporting compute virtual machine or in a containerisation platform. Due to the immutability of blockchain storage, a smart contract solution includes the use of a unique programming language. Furthermore, traditional software development lifecycles, especially regarding update and release cycles, are no longer considered valid approaches \cite{Destefanis2018}. A one-time release of a smart contract is required, therefore, the accuracy of the software development and testing is paramount. Due to this, novel development considerations need to be considered when implementing smart contracts, moreover, smart contracts and the associated programming language are blockchain specific. 

The Ethereum blockchain was explicitly designed to support smart contracts, with the architecture being turing complete, meaning it can represent any algorithm \cite{Teller1994} \cite{Hopcroft2001}. Aspects of smart contracts on this blockchain will be explored beginning with the compute cycles. These are accounted for, described as the execution time of a smart contract, and are limited to a variable called ‘gas’, paid for with Ether, Ethereum’s cryptocurrency. The result of exceeding this variable is the premature termination in computation of the smart contract \cite{li2020}. Smart contract code execution needs to be tested to ensure it remains within these limits. In addition to development correctness, all aspects of the contract need to be finalised, such as the wallet address of each party. Unlike traditional contracts, there is no option for amendments. 

Smart contracts are not limited to party transactions and can be linked together to perform a greater task. This has brought about the realisation of distributed applications, applications within a replicated blockchain \cite{Tonelli2018}. These applications function as the back-end and provide an \emph{Application Programming Interface}  (API), with the front-end being developed to use this API to produce the presentation of an interface for the application. This can be in the form of a web interface \cite{gao2019}. Smart contract development has evolved to support different programming languages, and therefore different syntax. Smart contract structure remains application specific, but is a recognised data type at the data type layer, in figure \ref{fig:blockchainstoragecomponents}.

\subsubsection{Transactions}
\label{transactionledger}

From our blockchain model, in figure \ref{fig:blockchainstoragecomponents}, we next discuss the transaction data type. Transactions can represent updates to any data structure existing at the organisation layer, with the structure of an update remaining application specific. Transactions have commonly been associated with and organised to form a ledger, with the transaction being used to modify ownership of items existing in a distributed ledger. This can either represent a physical or tangible item as a token, often called a non-fungible token (NFT) or the ownership of a digital asset such as a cryptocurrency. A single transaction can be between multiple parties, it a Bitcoin blockchain transaction this is expressed as multiple outputs in the transaction definition \cite{gobel2017}. Transactions have also been used to update other objects stored on a blockchain that are not in the organisation of a ledger. Such objects include medical records \cite{kuo2017} and the storage of IoT output data \cite{guin2018}. The structure of transactions remains application-specific; at present, it is the most common data structure in use. This exists at the data type layer in figure \ref{fig:blockchainstoragecomponents}.

\subsubsection{File Storage}
\label{blockchainfilestorage}

 The file storage component existing at the data type layer, in figure \ref{fig:blockchainstoragecomponents}, is explored next. Blockchains implemented for a file storage solution often do not store the file on the blockchain, but store a representation of a file in the form of a hash digest or only the associated metadata belonging to the file, specifically its location on external storage \cite{Cui2018} \cite{Shafagh2017}. However, with an appropriate hashing algorithm and the consideration of the network size, that is the number of replicas of a blockchain, therefore, the total storage consumed, files can be stored within blocks on a blockchain \cite{Dai2018}. A blockchain supporting small files, such as text documents, transforms them into an immutable document without the need for external encoding. A similar solution was purposed for the storage of system log files \cite{Pourmajidi2018}. Direct file storage remains primarily theoretical. However, it is a consideration for a data type at the data type layer, displayed in our novel model in figure \ref{fig:blockchainstoragecomponents}.

\subsubsection{Objects and Other Data Packages}
\label{objectsandotherdatapackages}
 Objects and other data packages have been acknowledged as a data type at the data type layer in figure \ref{fig:blockchainstoragecomponents}. There are currently active blockchain based applications used to store message communications, comments to published media, and blog content \cite{chowdhury2020}. Other message based objects can be stored on a blockchain as data structures, such as JavaScript object notation (JSON), extensible markup language (XML) and YAML. Each of these natively manages the organisation of data into information through the use of a key structure. Such structures are universally accepted and are designed for interoperability between technology solutions \cite{maeda2012}. This grouping has significant research and experimental potential within blockchain based solutions, and is used as a placeholder for future developments.

\section{Future Considerations}
\label{futureconsiderations}

Application-specific implementations of blockchain and its associated technologies continue to be the dominant approach to blockchain storage solutions. In recent years, blockchain and distributed ledger projects have been growing to support reusable, open-source code, in a modular design. Modular means the ability to add components from the storage layer presented in figure \ref{fig:blockchaincomponentslayers} and discussed in section \ref{blockchaincomponetsdefined}. This removes the necessity to design a blockchain in its entirety and therefore an application-specific solution, as not all blockchains are suited to every solution. These blockchain projects focus on generalised implementations that suit a range of solutions but differentiate between each project by highlighting the individual strengths, making it more likely to find an out-of-the-box blockchain solution. In figure \ref{fig:blockchainstoragecomponents}, blockchain and its associated components are identified and generalised, however, due to current application-specific implementations and the lack of detail into how the solution uses blockchain, as an immutable storage technology, this model is incomplete. Placeholders of ‘other’, at the organisation and data type layer, allow for the extension of this model as a future consideration to extend the components as each type is standardised.

Standardisation will benefit blockchain’s adoption, the interoperability of blockchain solutions, and the development practices that extend into all aspects of blockchain; being its associated technologies, components and terminologies. Future considerations include the introduction of \emph {Request For Comment} (RFC) like documentation or an ‘International Organization for Standardization’ guidance to outline the specification for each of the technology components, to assist in the compatibility of immutable storage solutions both in the design and development phases. In is acknowledged that work has been started in a formalisation process within the Institute of Electrical and Electronics Engineers (IEEE), presented in \cite{IEEE2020}. It is envisioned that such an approach will result in an alignment between the theoretical understanding and the practical application of blockchain as an immutable storage solution and any associated blockchain technologies.

\section{Conclusion}
\label{conclusion}
This paper presents a novel layered model to categorise technologies and terminologies that are either directly required for a blockchain's core function at the base layer and the associated optional components in the layers above this. Article presents a universally recognised logical structure that can be used as a guide when referencing and building a blockchain storage solution. This helps to address the current lack of consistency in the implementation and interpretation of blockchain architecture as emphasised by a review of a selected literature on the topic. Our proposed layered model produces a modular structure to allow for a granular selection of blockchain components comprised from all known blockchain technologies. Each categorisation is explored in detail to build a group of related concepts and technologies. By generalising blockchain as an immutable online storage solution, we highlight the key aspects that have made it a valuable technology making it the only method for continuous write-once, read-only online storage and achieving this without governance.

 \nocite{*} 
\bibliographystyle{unsrt}
\bibliography{IEEEbib}  






\end{document}